\documentstyle[12pt]{article}
\newcommand{\beqar}{\begin{eqnarray}}
\newcommand{\eeqar}[1]{\label{#1} \end{eqnarray}}

\def\slash#1{\not\!#1}

\def\picture #1 by #2 (#3 scaled #4)  
  {\dimen0=#1  \dimen1=#2             
   \divide\dimen0 by 1000             
   \multiply\dimen0 by #4             
   \divide\dimen1 by 1000             
   \multiply\dimen1 by #4
   \vbox to \dimen1
     {\hrule width \dimen0 height 0pt depth 0pt
      \vfill
      \special{picture #3 scaled #4}
     }
  }

\def\Ref#1{$^{#1}$}

\begin{document}

\thispagestyle{empty}  

\rightline{UTAPHY-HEP-4}
\rightline{September 10, 1992}

\vskip 0.5in
\begin{center}
  {\large \bf DOUBLE BETA DECAY}
  \vskip 4ex
  S. P. Rosen
  \vskip 1ex
  {\it College of Science \\
       The University of Texas at Arlington\\
       Arlington, Texas 76019}
\end{center}

\vskip 4ex
\centerline{ABSTRACT}
\begin{quotation}
\indent
I review the subject of double beta decay with emphasis upon its implications
for the neutrino. It is shown that even if right-handed currents provide the
phenomenological mechanism for no-neutrino decay, the fundamental mechanism
underlying the process must be neutrino mass. Estimates of the mass required
suggest that this mechanism is less likely than the direct mechanism of
Majorana mass terms in the mass matrix.
\end{quotation}

\vskip 0.5in
\begin{center}\sl
Invited talk at the Franklin Symposium honoring \\
Frederick Reines and the discovery of the neutrino \\
held at the Franklin Institute, Philadelphia \\
April 30, 1992.
\end{center}

\vfil

\begin{quote}
\small This research is supported in part by the U. S. Department of
Energy under grant DE-FG05-92ER40691.
\end{quote}

\textheight=8.5in
\newpage
\setcounter{page}{1}

\begin{center}
\begin{Large}
{\bf
DOUBLE BETA DECAY
}\\
\end{Large}
\vskip 5ex

             S. P. Rosen\\
\vskip 1ex
{\it        College of Science\\
            The University of Texas at Arlington\\
            Arlington, Texas 76019\\
} 
\end{center}

\section{Introduction}
Nature has a way of teasing us with her most intimate secrets: she wants us to
know that they are there, but she is reluctant to let us find out exactly what
they are. So she throws out a clue to catch our attention and, once we have
found it, she puts all kinds of obstacles in our way and draws false leads
across our path. It is only with great persistence and painstaking attention to
every detail that we have any chance of wresting some of these secrets from
her.

Fred Reines is a physicist who loves to take up such challenges. Just as he
set out to observe the neutrino, which had eluded us for twenty five years, and
was prepared to explode a bomb to do it,\Ref{1} so he has gone seeking for
proton decay, for supernovae, for neutrino-electron interactions, and for
double beta decay. All are either rare, or extremely rare phenomena, but all
are
associated with some fundamental principle of physics; and it is this
association that adds so much spice to the pursuit.

Inevitably the pursuit takes a long time. In the case of double beta decay,
it took thirty years to show that the phenomenon does indeed occur in ancient
ores\Ref{2} and another twenty to observe an actual double beta event in the
laboratory.\Ref{3} In the mid fifties, Fred and his colleagues Cowan, Harrison,
and  Langer\Ref{4} undertook a pioneering search for $^{150}\hbox{Nd} \to
^{150}$Sm, which  has a relatively large $Q$-value of 3.5 MeV and which Mike
Moe
has only recently  started to re-investigate. Today we are still searching for
one form of double  beta decay, which, should it occur, will have profound
implications for the  question of neutrino mass and for physics beyond the
standard electroweak model. The  problem is, and always has been that double
beta decay is so slow that minute  traces of intrinsically faster
radioactivities, such as those associated with  the Uranium and Thorium decay
chains, become significant backgrounds mimicking  the signal for double beta
decay.

Like its sister phenomenon of single beta decay, double beta
decay is a process in which a nucleus undergoes a transmutation from one
element to another. Whereas in single beta decay, the atomic number usually
increases by one unit, from $Z$ to $Z+1$, and an electron, or negative beta
ray, is emitted in order to conserve electric charge, so in double beta decay
$Z$ increases by two units and two electrons are emitted. Lifetimes of single
beta decay vary from fractions of a second to hundreds of thousands of years
depending upon the nature of the nuclear transmutation and the total amount of
energy released in the process. Lifetimes of double beta decay are tens of
billions of billions of years and longer.

To ensure that energy, momentum, and angular momentum are conserved
in single beta decay, each beta ray is always accompanied by a neutral particle
of spin $1/2$; by convention the particle accompanying a negative beta ray is
called an anti-neutrino and the one accompanying a positive beta ray is called
a neutrino. Since negative and positive beta rays are themselves particle and
anti-particle respectively, namely electrons and positrons, the nomenclature is
clearly designed to suggest another conservation law. Assigning a leptonic
`charge' of $+1$ unit to `leptons', that is, electrons and neutrinos, and $-1$
to anti-leptons, namely positrons and anti-neutrinos, we might be tempted to
think that lepton charge, or lepton number is conserved in beta decay. But is
that really the case?

According to the standard model of Glashow, Weinberg, and Salam,\Ref{5} the
neutrino is distinguished from its anti-particle by virtue of helicity: the
neutrino is perfectly left-handed and the anti-neutrino perfectly right-handed.
As long as the neutrino has precisely zero mass, these two particle states are
orthogonal to one another and cannot  `communicate'; but if the neutrino
acquires a mass, no matter how small, then the possibility arises that neutrino
and anti-neutrino will no longer be orthogonal. Such a neutrino is called a
Majorana particle\Ref{6} and there are numerous theories beyond the standard
model,  for example those unifying all the known interactions, which predict
that the  neutrino is indeed a Majorana particle.

The key test for a Majorana neutrino, originally proposed by G.Racah\Ref{7} in
1937, is whether an anti-neutrino emitted in the beta decay of one neutron can
interact with another neutron and cause it to transform into a proton and an
electron. Such a combination of events would violate the conservation of lepton
number because there would be two neutrons and no leptons in the initial state
and two protons, two electrons and no anti-neutrinos in the final state; lepton
number would have changed from zero to 2 units. With the discovery of  parity
nonconservation and the two-component neutrino in 1957, it was  recognised that
the two-step process of Racah is inhibited by helicity: the  right-handed
anti-neutrino emitted by the first neutron is in the wrong  helicity state to
be
re-absorbed by another neutron. In order to complete the second step of the
Racah process, the anti-neutrino must be able to flip its  helicity and turn
itself into a neutrino. We show below that such a double flip may be induced by
a mass term unique to electrically neutral fermions and known as a Majorana
mass
term.

One way in which to study the Racah process is to use real anti-neutrinos.
This was, in fact the first experiment Ray Davis\Ref{8} performed with the
famous  $^{37}$Cl $\to$ $^{37}$Ar reaction, using anti-neutrinos from an atomic
reactor  in place of neutrinos from the sun.
(Rumors of a positive result reached Bruno Pontecorvo\Ref{9} in Moscow in 1957
and  caused him to invent neutrino oscillations in direct analogy with the
Gell-Mann--Pais analysis of neutral Kaon decay. The rumors eventually died out
but the idea of oscillations is still alive and kicking.) A much more sensitive
method is to study double beta decay.

Double beta decay, because of its extremely long lifetime, is regarded as a
second-order effect of the same interaction as gives rise in first order to
single beta decay. Neutrinos are not needed to  conserve energy and momentum in
double beta decay and so we may ask whether  they must always accompany the
emitted electrons. If the neutrino and anti-neutrino are Majorana particles and
hence not orthogonal to one another, then it is possible to reproduce the Racah
process inside the nucleus: the anti-neutrino emitted by one neutron may be
reabsorbed by a second to give a form of  decay in which no neutrinos
materialise. W. H. Furry,\Ref{10} in 1939, was the first to  realise this
possibility, and he showed that, absent the helicity suppression,  the virtual
neutrino was much better at stimulating the Racah process because  its
effective
energy would be much greater than that of a real neutrino. The  enhancement of
the rate could be as large as six orders of magnitude and this  makes double
beta decay sensitive to very small lepton number nonconserving  effects.

We see from this analysis that there are two forms of double beta decay, the
two-neutrino decay in which two neutrinos are emitted together with the two
electrons and the no-neutrino decay in which no neutrinos are emitted.
Two-neutrino decay will occur irrespective of the nature of the neutrino, but
no-neutrino decay can occur only if the neutrino is a Majorana particle. It is
also worth mentioning that certain theories of neutrino mass predict other
forms of decay in which no neutrinos are emitted, but the two electrons are
accompanied by one or more scalar particles called Majorons.\Ref{11} We shall
argue  below that the occurrence of no-neutrino decay implies that the neutrino
must  have a non-zero mass and sets lower bounds on this mass.

Before delving more deeply into the theory of the neutrino, let us consider the
nuclear setting in which double beta decay takes place and the original clue
that Nature put in our way. The story began in the 1930's with the
question of the stability over geological time of certain nuclei which, on the
basis of energetics alone, could not be absolutely stable. These nuclei are all
composed of an even number of neutrons $N$ and an even number of protons $Z$
and, because the nuclear force tends to bind pairs of like particles more
tightly than pairs of different particles, they are lighter than neighboring
odd-odd nuclei which contain the same total number of nucleons $A = N + Z$, but
odd numbers of neutrons and of protons. It can, and does happen that a
particular even-even nucleus $(A,~N,~Z)$ is lighter than its nearest neighbor
$(A,~N-1,~Z+1)$ but heavier than the next nearest neighbor $(A,~N-2,~Z+2)$.
Therefore, while decay from $Z$ to $Z+1$ is forbidden, decay from $Z$ to $Z+2$
is allowed energetically and it should indeed take place.

The problem is to find a decay mechanism sufficiently slow that known even-even
nuclei, for example $^{130}$Te and $^{82}$Se could survive in significant
quantities over periods of the order of several billion years. In 1935 Maria
Mayer\Ref{12} realised that second-order beta decay could provide the needed
mechanism. She used the then new Fermi theory of beta decay to estimate the
lifetime for  the two-neutrino mode, and found it well in excess of $10^{17}$
years, which is exceedingly slow even on geological time scales. Four years
later Wendell Furry\Ref{10} estimated the lifetime of no-neutrino decay to be a
million times shorter than  the two-neutrino one; nevertheless it was still
long
enough to account for the  stability of even-even nuclei.

Today we know that the lifetimes of both modes are much longer than these early
estimates. Two-neutrino decay has been observed in several nuclei with
lifetimes ranging from $10^{19}$ years in $^{100}$Mo to $10^{21}$ years in
$^{130}$Te. Prolonged searches for no-neutrino decay have been made, so far
without success, and the best lower bound on its lifetime is of order $10^{24}$
years in the decay of $^{76}$Ge. In this talk I shall discuss the implications
of these results and show how they may be used to set lower bounds on the mass
of neutrinos.

\section{The neutrino mass matrix}

I now turn to neutrino properties. In the context of the Dirac equation for
spin $1/2$ particles, the mass term serves to change helicity from left to
right and vice versa. Consider for the  moment an electron: its four states can
be described as an electron with left-handed helicity (spin anti-parallel to
momentum) $e^-_L$, an electron with  right-handed helicity (spin parallel to
momentum) $e^-_L$, a left-handed  positron $e^+_L$, and a right-handed positron
$e^+_R$. The mass term  transforms the two particle states into each other
\beqar
                    e^-_L \leftrightarrow e^-_R~,
\eeqar{1}
and the two anti-particle states into each other
\beqar
                    e^+_L \leftrightarrow e^+_R~,
\eeqar{2}
but the transformation of a left-handed particle into a right-handed
anti-particle
\beqar
                    e^-_L \leftrightarrow e^+_R
\eeqar{3}
is forbidden by the conservation of electric charge.

Next consider the four corresponding electron-neutrino states: a left-handed
neutrino $\nu_{eL}$, a right-handed neutrino $\nu_{eR}$, a left-handed
anti-neutrino $\bar{\nu}_{eL}$, and a right-handed anti-neutrino
$\bar{\nu}_{eR}$. In the standard model only $\nu_{eL}$ and $\bar{\nu}_{eR}$
take part in weak interactions. Exactly in parallel with the electron mass
term, a neutrino mass term could induce left-right transitions between
particles and between anti-particles
\beqar
                  \nu_{eL} &\leftrightarrow & \nu_{eR} \nonumber \\
                  \bar{\nu}_{eL} &\leftrightarrow & \bar{\nu}_{eR}~,
\eeqar{4}
but in addition electric charge conservation no longer forbids transitions
between particle and anti-particle
\beqar
                  \nu_{eL} &\leftrightarrow & \bar{\nu}_{eR} \nonumber \\
                  \bar{\nu}_{eL} &\leftrightarrow & \nu_{eR}~.
\eeqar{5}
These latter transitions are exactly what we need to make the Racah process
work and the presence of such terms in the neutrino mass-matrix gives rise to
Majorana mass eigenstates.

To see how this comes about in a formal way, let us consider the
electron-neutrino field $\psi$ and its anti-neutrino field
$\psi^c = C\bar{\psi}^T \equiv \gamma_2 \psi^*$.  The helicity projections of
the field are
\beqar
   \psi_L = \frac{1}{2}(1+\gamma_5)\psi; \hspace{3em}  \psi_R = \frac{1}{2}(1-
              \gamma_5)\psi,
\eeqar{6}
and those of the charge conjugate are
\beqar
(\psi_L)^c = \frac{1}{2}(1-\gamma_5)\psi^c =(\psi^c)_R \nonumber \\
(\psi_R)^c = \frac{1}{2}(1+\gamma_5)\psi^c =(\psi^c)_L.
\eeqar{7}
The most general form of the mass matrix can be written in terms of all four
components of the fermion field as\Ref{13}
\beqar
  \matrix{(\, \bar{\psi}_L, & (\bar{\psi}_R)^c, &
          (\bar{\psi}_L)^c, & \bar{\psi}_R \,) \cr
          & & & \cr &  & & \cr & & & \cr}
  \pmatrix{ O & \pmatrix{m_L & m_D \cr m_D & m_R \cr} \cr \noalign{\smallskip}
            \pmatrix{m_L & m_D \cr m_D & m_R \cr} & O \cr}
  \pmatrix{ \psi_L  \cr (\psi_R)^c \cr (\psi_L)^c \cr \psi_R \cr}
\eeqar{8}
The matrix element $m_D$ gives rise to the transitions of eq.~(4) and is called
the Dirac mass term, while the matrix elements $m_L$ and $m_R$ give rise to the
transitions of eq.~(5) and are known as the Majorana mass terms.

We use the phrase `Majorana mass terms' because as long as one of $m_L$ and
$m_R$ is non-zero, the eigenvectors of the mass matrix are linear combinations
of  $\psi$ and $\psi^c$ and transform into themselves under the charge
conjugation operation. For example, if $CP$ is conserved and $m_L = m_R = m_M$,
then the mass eigenvalues are
\beqar
 M_{\pm} = \frac{1}{2}(m_M \pm m_D)
\eeqar{9}
and the corresponding eigenvectors are
\beqar
\Psi_{\pm} = \frac{1}{\sqrt2}(\psi \pm \psi^c)
\eeqar{10}
More generally, if we define
\beqar
L   & = & \pmatrix{\psi_L     \cr
                   (\psi_R)^c \cr} \nonumber \\
L^c & = & \pmatrix{(\psi_L)^c \cr
                   \psi_R     \cr}
\eeqar{11}
and $m_L \neq m_R$, then the eigenvectors of the mass matrix are
\beqar
L_{\pm} = \frac{1}{\sqrt2}(L \pm L^c)
\eeqar{12}

Although these fields have the formal appearance of eigenvectors of the
charge-conjugation operation $\psi \rightarrow \psi^c$ with eigenvalues $\pm
1$,
the  physical helicity states upon which they act are eigenstates of $CP$. This
is as it should be since the mass matrix conserves $CP$, but not $C$. We refer
to this eigenvalue as the $CP$ eigenvalue, noting that in the most general case
it is a  complex phase and that the parity of a Majorana field is actually
imaginary.  For our purposes, however, it is the opposite algebraic signs that
carry the  important physical information. We also note that the original
(Dirac) field  $\psi$ can be expressed as a linear combination of two Majorana
fields with  opposite $CP$ eigenvalues, for example
\beqar
\psi_{\hbox{\scriptsize Dirac}} = \frac{1}{\sqrt2}(\Psi_+ + \Psi_-)
\eeqar{13}

\section{Physical processes}

In the standard model, the effective beta decay Hamiltonian consists of
products of left-handed currents for nucleons and leptons with the appropriate
Hermitian adjoint currents. It gives rise to a set
of basic physical processes involving transitions between neutrons and
protons and the corresponding emission or absorption of leptons and
anti-leptons. We summarise these processes in the following Table.
\begin{quote}
\small
{\bf Table 1.}  Emission and Absorption of Neutrinos in the Standard
Model. The helicity configurations of leptons and anti-leptons are
denoted by the subscripts $R$ and $L$.
\end{quote}
\begin{center}
\renewcommand{\arraystretch}{1.5}
\begin{tabular}{|c|c|c|c|} \hline
Process    & Nucleon   & Beta    & Neutrino         \\
           & Transition& Ray     &                  \\ \hline
Emission   & $n \to p$ & $e^-_L$ & $\bar{\nu}_{eR}$ \\
Emission   & $p \to n$ & $e^+_R$ & $\nu_{eL}$       \\
Absorption & $n \to p$ & $e^-_L$ & $\nu_{eL}$       \\
Absorption & $p \to n$ & $e^+_R$ & $\bar{\nu}_{eR}$ \\ \hline
\end{tabular}
\end{center}
\vskip 1ex

How does this affect no-neutrino double beta decay? The first and third lines
of Table 1 illustrate the point made in the {\bf Introduction}, namely that in
order for the neutrino emitted in the first step of the process to be
reabsorbed in the second step, a
$\bar{\nu}_{eR}$ must be able to transform itself into a $\nu_{eL}$. One way to
accomplish this transformation is through the Majorana mass terms of eqs.~(5,
8)  above; another way is to include small admixtures of right-handed currents
in  the predominantly left-handed currents of the effective Hamiltonian,
treating  the electron-neutrino as a Majorana particle identical with its
anti-particle.  We can represent these admixtures insofar as neutrinos are
concerned by  modifying the first and third lines of Table 1:
\begin{quote}
\small
{\bf Table 2.}  Emission and Absorption of Neutrinos in a Modified
Standard  Model. The parameter $\eta$ denotes small admixtures of opposite
helicities and  the assumption of a Majorana neutrino means that $\bar{\nu}
\equiv \lambda \nu$  where $\lambda = \pm 1$.
\end{quote}
\begin{center}
\renewcommand{\arraystretch}{1.5}
\begin{tabular}{|c|c|c|c|} \hline
Process    & Nucleon   & Beta  & Neutrino \\
           & Transition& Ray   &          \\ \hline
Emission   & $n \to p$ & $e^-$ & $\bar{\nu}_{eR} + \eta \bar{\nu}_{eL}$ \\
Absorption & $n \to p$ & $e^-$ & $\nu_{eL} + \eta \nu_{eR}$ \\ \hline
\end{tabular}
\end{center}
\vskip 1ex

The leptonic part of the no-neutrino double beta decay is thus proportional to:
\beqar
{\cal L} & = &\frac{1}{\sqrt2}(1 - P(e_1,e_2))
    \langle L_{\mu}L_{\alpha} \rangle \nonumber \\
\langle L_{\mu}L_{\alpha} \rangle & = & [\overline{e^-_1} \gamma_{\mu}
    \overbrace{(\nu_L + \eta \nu_R)]\,[(\nu_L + \eta \nu_R)^T}
    \gamma_{\alpha}^T (\overline{e^-_2})^T]~,
\eeqar{14}
where the overbrace indicates that we must contract the neutrino fields using
the propagator
\beqar
\frac{i\slash{q} + m}{q^2 + m^2}~.
\eeqar{15}
Fields with the same helicity are linked by the mass term and those with
opposite helicity by the $\slash{q}$ term.  Thus the two mechanisms have
amplitudes proportional to:
\beqar
\hbox{Mass-mechanism} &\sim & \frac{\lambda m}{q^2 + m^2} \nonumber \\
\hbox{RHC-mechanism}  &\sim & \frac{\lambda \eta \slash{q}}{q^2 + m^2},
\eeqar{16}
where $\lambda$ is the $CP$ eigenvalue of the neutrino.

Although we are treating the admixture of right-handed currents (RHC-mechanism)
as a separate {\em phenomenological} mechanism for no neutrino double beta
decay, we intend to show later that it is not a separate {\em fundamental}
mechanism. We shall argue that in gauge theories, the mass is the fundamental
mechanism for lepton number nonconserving processes. In other words the
RHC-mechanism will not work unless there are mass terms present in the
neutrino mass matrix.

\section{Neutrino mixing}

Although the preceding discussion has treated the electron-neutrino as if it
were a single particle, we must recognise that, like its companion muon- and
tau-neutrinos, it is really a superposition of the mass eigenstates of the
neutrino mass matrix. To allow for the three families, each of the entries in
the row and column of fields in eq.~(8) becomes a 3-fold object and each entry
in the matrix becomes a 3-by-3 matrix ; in addition, the sub-matrix in the
lower left quarter becomes the Hermitian adjoint of the one in the upper right
quarter. As long as the Majorana matrices corresponding to $m_L$ and $m_R$ do
not vanish, the eigenvectors will be CP eigenstates analogous to those in
eqs.~(10, 11, 12) and with eigenvalues $\pm1$. We denote them as $\nu_i$ with
mass  $m_i$ and CP eigenvalue $\lambda_i$:
\beqar
 (\nu_i)^c = \lambda_i \nu_i~,  \hspace{5em} \lambda_i = \pm 1~.
\eeqar{17}

The flavor eigenstates $\nu_{eL}$ and $\nu_{eR}$ can now be written as linear
superpositions of $\nu_{iL}$ and $\nu_{iR}$:
\beqar
\nu_{eL} &= & \sum_i U_{ei} \nu_{iL} \nonumber \\
\nu_{eR} &= & \sum_i V_{ei} \nu_{iR} ,
\eeqar{18}
and similarly for $\nu_{\mu}$ and $\nu_{\tau}$. The mixing matrices $U_{ei}$
and $V_{ei}$ are both unitary,
\beqar
\sum_i U_{ei}^* U_{ei} = \sum_i V_{ei}^* V_{ei} = 1
\eeqar{19}
but not necessarily orthogonal to one another.

The leptonic factors for the two mechanisms for no-neutrino double beta decay
described in eq.~(16) must now be modified to take neutrino mixing into
account.
For the mass mechanism, we replace the expression in eq.~(16) by an effective
mass $m_{\beta\beta}$, where
\beqar
m_{\beta\beta} = \sum_i m_i \lambda_i (U_{ei})^2;
\eeqar{20}
and for the RHC-mechanism we use
\beqar
\eta_{LR} = \sum_i \lambda_i U_{ei}V_{ei} \frac{\slash{q}}{q^2 + (m_i)^2}~.
\eeqar{21}
The effective mass $m_{\beta\beta}$ obviously vanishes when all the mass
eigenvalues $m_i$ vanish, but $\eta_{LR}$ would appear not to do so. In other
words, it would seem that the RHC-mechanism could give rise to no-neutrino
double beta decay even in the absence of neutrino mass. We shall now argue that
this cannot happen in gauge theories because of their renormalizability.

\section{Limitation from high energy behavior}

The renormalizability properties of gauge theories implies that the amplitudes
for physical processes cannot diverge at high energies: the amplitudes for
specific diagrams contributing to such processes will either be finite
themselves at high energy or their divergent parts will cancel amongst one
another. For example, the creation of a W-boson pair in $e^+ e^-$ annihilation
at high energies can take place through two diagrams:
$$
  \picture 7.42in by 1.81in (Fig1 scaled 500)
$$
\begin{quote}
\small
{\bf Figure 1.}~~Feynman diagrams for $e^+e^- \to W^+W^-$ via neutrino and
$Z^0$
exchange.
\end{quote}
one involving neutrino exchange and the other a virtual $Z^0$-boson. By
themselves, these  diagrams are quadratically divergent at high energies; but,
because of the relationships amongst coupling constants imposed by the standard
Glashow-Weinberg-Salam model, the divergent parts of the two diagrams cancel
each  other.

If we look carefully at the diagram for no-neutrino decay, we find a
sub-diagram which corresponds to the process (Fig.~2)
\beqar
W^- W^- \to e^- e^- .
\eeqar{22}
The amplitude via the mass mechanism does not diverge at high energies, but the
amplitude via the RHC-mechanism does. Gauge theories must therefore produce a
way of canceling this divergence. One way would be to have a gauge group with
doubly-charged gauge bosons to cancel the neutrino exchange of Fig.~2. There
is, however, no evidence for the existence of such bosons, and so we must
confine ourselves to models without them. In this case the only way to
eliminate the divergent part of the amplitude is to require its coefficient to
be zero.
$$
  \picture 5.24in by 2.97in (Fig2 scaled 500)
$$
\begin{quote}
\small
{\bf Figure 2.}~~Feynman diagrams for no-neutrino double beta decay. Note the
subdiagram corresponding to  $W^- W^- \to e^- e^-$.
\end{quote}

It is not difficult to show that this coefficient is:
\beqar
A_{LR} \equiv \sum_i \lambda_i U_{ei}V_{ei}.
\eeqar{23}
Therefore the gauge theory requirement of good high energy behavior for
$W^- W^- \rightarrow e^- e^-$ is that
\beqar
A_{LR} = 0
\eeqar{24}
This is a completely general requirement and, as shown by Kayser, Petcov, and
Rosen,\Ref{14} it holds in a large class of gauge theories. Comparing eqs.~(23,
24) with eq.~(21) above, we see that the vanishing of $A_{LR}$ is sufficient to
forbid  no-neutrino double beta decay when all neutrino masses vanish, or when
they are  all completely degenerate. In the case when all the neutrino mass
eigenvalues  are small compared with $q^2$, we can expand eq.~(21) in powers of
$(m_i/q)^2$  to obtain
\beqar
\eta_{LR} = \sum_i \lambda_i U_{ei}V_{ei} \left(\frac{\slash{q}}{q^2}\right)
            \frac{- (m_i)^2}{q^2}.
\eeqar{25}
Note that the typical value of $q^2$ corresponds to a mean inter-nucleon
separation of a few fermi and has a value of order (50 MeV)$^2$.

\section{Nuclear physics}

Most double beta decay transitions take place between the groundstates of
even-even nuclei which always have zero spin and positive parity: $0^+ \to
0^+$.
Many daughter nuclei have excited states with spin and parity $2^+$ about 500
keV above the groundstate and transitions from the parent to these excited
states are also possible. In the case of no-neutrino decay, such $0^+ \to 2^+$
transitions can only arise through the RHC-mechanism\Ref{15} and so  the
observation of them would provide important information about the {\em
phenomenological} mechanism for double beta decay.

Typical energy releases ($Q$) in double beta decay are in the range of 2--3
MeV,
but there are examples like $^{128}$Te $\to$ $^{128}$Xe and $^{238}$U $\to$
$^{238}$Pu with $Q$-values closer to 1 MeV. Two-neutrino decay has a four-body
phase space corresponding to the two electrons and two neutrinos  emitted and
it
is usually a polynomial of degree 10-11 in $Q$:
\beqar
F_{2\nu} = \hbox{Phase Space} \sim Q^{10-11}
\eeqar{26}
No-neutrino decay has a two-body phase space together with an integration over
the energy of the virtual neutrino; the phase space is a polynomial of degree 5
in $Q$, and the integration is proportional to the fifth power of the energy
$E_{\nu}$ of the virtual neutrino. Thus the no-neutrino factor corresponding to
the four-body phase space of two-neutrino decay is:
\beqar
\Phi_{0\nu} = (E_{\nu})^5 F_{0\nu} \sim (E_{\nu}Q)^5
\eeqar{27}
For a typical separation between neutrons inside the nucleus of 4--5 fermi, the
mean energy of the virtual neutrino is about 50 MeV and so the ratio of phase
space factors favors the no-neutrino decay over the two-neutrino mode by a
factor
\beqar
R\left( \frac{0\nu}{2\nu} \right) \sim \left( \frac{E_{\nu}}{Q}\right)^5
                                   \sim 10^6.
\eeqar{28}
It is this factor that enables us to set sensitive limits on lepton number
nonconserving parameters.

\section{Nuclear matrix elements and two-neutrino half-lives}

In second order perturbation theory, the matrix element for double beta decay
takes the general form
\beqar
M(i \rightarrow f) = \sum_k \langle f\mid H_{\beta}\mid k \rangle
                     \frac{1}{E_k - E_i} \langle k\mid H_{\beta}\mid i \rangle
\eeqar{29}
where $i$ denotes the initial state of the parent nucleus, $k$ the intermediate
state of the nucleus plus one electron and one neutrino, and $f$ the final
state
of the nucleus plus two electrons and either two neutrinos or no neutrinos. The
energies of these states are denoted by $E_k$ and $E_i$. To carry out the sum
over intermediate states, we replace the energy denominator by an average value
and use closure over the states of the intermediate nucleus. While closure is a
good approximation for the no-neutrino mode, its validity for two-neutrino
decay is a matter of some debate.\Ref{16}

For $0^+ \rightarrow 0^+$ nuclear transitions the dominant contribution to the
nuclear matrix element comes from the axial vector part of the effective weak
Hamiltonian $H_{\beta}$. The matrix elements for two-neutrino and no-neutrino
decay are respectively:
\def\bfsigma{\hbox{\boldmath $\sigma$}}
\beqar
M^{2\nu} & = & \langle f \mid \sum_{jk} \tau_j \tau_k \,
               \bfsigma_j \cdot \bfsigma_k
               \mid i \rangle  \\ \nonumber
M^{0\nu}(m_n) & = & \langle f\mid \sum_{jk} \tau_j \tau_k \,
               \bfsigma_j \cdot \bfsigma_k
               \frac{I(r_{jk},m_n)}{r_{jk}} \mid i \rangle~.
\eeqar{30}
where each operator $\tau_l$ transforms a neutron in the initial state into a
proton in the final state and the function $I(r_{jk},m_n)/r_{jk}$ in $M_{0\nu}$
represents the propagator for a neutrino of mass $m_n$. When we multiply the
no-neutrino matrix element by the lepton number nonconserving parameters of
eqs.~(20, 21) we must carry out a sum over $n$.

The integration over the neutrino momentum in the function $I(r_{jk},m_n)$
leads to standard functions in the case of zero neutrino mass, but it is much
harder to carry out when the mass is nonzero.\Ref{17} Haxton\Ref{18} has found
an empirical  representation which works well for small masses. Expanding this
representation  to lowest order in $m_n$, we find that
\beqar
\frac{I(r,m)}{r} \approx \frac{I(r,0)}{r}
                 - \frac{m^2}{a \langle E_{ki} \rangle}~,
\eeqar{31}
where a is a slowly varying parameter roughly equal to 0.4 in the range of
interest, and $\langle E_{ki} \rangle$ is an average value of the energy
denominator of eq.~(29).

That the lowest term in eq.~(31) is quadratic in $m$ is easily understood from
the  propagator in eq.~(16). It is also noteworthy that this term is
independent
of  the nucleon separation variable $r$, just like the two-neutrino matrix
element $M^{2\nu}$ of eq.~(30). This means that the corresponding contribution
to the  no-neutrino lifetime can be expressed in terms of the two-neutrino
lifetime,\Ref{18} a  result which will be very useful when the term independent
of $m$ vanishes.

We can now write the half-life for two-neutrino decay in the form\Ref{17}
\beqar
\frac{1}{\tau^{2\nu}} = F_{2\nu}(Q) |M^{2\nu}|^2 \langle E_{ki}\rangle^{-2}~,
\eeqar{32}
where $\langle E_{ki} \rangle$,the average energy denominator in eq.~(29),is
roughly one-half the energy release $Q$. The matrix elements and resulting
half-lives for several nuclei have been calculated by various authors using the
shell model and using the quasi-particle random phase approximation (QRPA); the
results\Ref{19} and the comparison with experiment are given in the following
Table 3.

\begin{quote}
\small
{\bf Table 3.} Calculated and Measured Half-Lives for Two-Neutrino Double Beta
Decay
\end{quote}
\begin{center}
\renewcommand{\arraystretch}{1.5}
\begin{tabular}{|c|c|c|c|c|c@{~Ref.\,}l|} \hline
  Parent  & $Q$-Value & $F_{2\nu}$
               & \multicolumn{4}{c|}{$\tau^{2\nu}$~~($10^{20}$yr)}\\
\cline{4-7}
 Nucleus  & (MeV) & (MeV$^2/10^{21}$yr) & Shell Model & QRPA
                                          & \multicolumn{2}{c|}{Expt.} \\
\hline
$^{76}$Ge & 2.041 & 34.70& 10.1&  16--63  & $9\pm1$                &20 \\
\hline
$^{82}$Se & 2.995 & 1151 & 0.64&   1--6   & $1.1^{+0.8}_{-0.3}$    & 3 \\
\hline
$^{100}$Mo& 3.034 & 2502 & --- &0.4--0.04 &$0.12^{+0.034}_{-0.01}$ &21 \\
\hline
\end{tabular}
\end{center}
\vskip 1ex

\noindent The comparison between theory and experiments works quite well.

\section{No-neutrino half-lives and neutrino mass bounds}

In this section we give a somewhat simplified version of the no-neutrino
half-life, keeping the leading matrix elements and dropping smaller ones. For a
more complete account the reader is referred to the recent review article by
Tomoda.\Ref{17} The matrix elements of eqs.~(30, 31) are functions of the mass
of
the exchanged  neutrinos and so we must sum them over the spectrum of mass
eigenstates using  weighting factors appropriate to the mass- and
RHC-mechanism.
The half-life can then be written as
\beqar
     \frac{1}{\tau^{0\nu}} = A_{mm} + 2B_{m\eta} + C_{\eta \eta},
\eeqar{33}
where $A$ arises from the mass-mechanism, $C$ from the RHC-mechanism and $B$
from the interference between them:
\beqar
 A_{mm} &= & \left[ \sum_n m_n \lambda_n (U_{en})^2 M^{0\nu}(m_n) \right]^2
                                                         F_{11}^0 \nonumber \\
 B_{m\eta} &= & \left[\sum_n m_n \lambda_n (U_{en})^2 M^{0\nu}(m_n)\right]
      \left[\sum_n \lambda_n U_{en}V_{en}M^{0\nu}(m_n) \right] 2F_{13}^0
                                                                  \nonumber \\
 C_{\eta \eta} &= & \left[\sum_n \lambda_n U_{en}V_{en}M^{0\nu}(m_n) \right]^2
                                                                  4F_{33}^0
\eeqar{34}
The $F_{kl}^0$ are phase space and Coulomb factors defined and tabulated by
Tomoda.\Ref{17} In the RHC-mechanism part of these expressions we have kept
only
those  terms which will give a significant contribution when the gauge theory
condition $A_{LR} = 0$ of eqs.~(23, 24) is satisfied.

Using the gauge theory condition and the expansion of eq.~(31) for the neutrino
propagator, we can rewrite the $A,~B,~C$ expressions of eq.~(34) as
\beqar
 A_{mm} &= & \left[m_{\beta \beta} M^{0\nu}(0) \right]^2 F_{11}^0 \nonumber \\
 B_{m\eta} & = & \left[m_{\beta \beta} M^{0\nu}(0) \right]
           \left[\sum_n \lambda_n U_{en}V_{en}\frac{(m_n)^2}
           {a \langle E_{Ni} \rangle} M^{2\nu}\right] 2F_{13}^0 \nonumber \\
C_{\eta\eta}& = & \left[ \sum_n \lambda_n U_{en}V_{en}
           \frac{(m_n)^2}{a\langle E_{Ni} \rangle}M^{2\nu} \right]^24F_{33}^0~.
\eeqar{35}
Notice that the two-neutrino matrix element now appears in the
coefficients $B$ and $C$ because of the absence of the nucleon separation
variable in the second  term of eq.~(31).

At the present time, the best limit on no-neutrino decay comes from studies of
$^{76}$Ge $\to$ $^{76}$Se which show that the half-life must be longer than
about $2\times10^{24}$ yrs.\Ref{22}  In terms of an effective mass for double
beta decay  this gives a limit of 1 eV:\Ref{17}
\beqar
m_{\beta\beta} \equiv \sum_n m_n \lambda_n (U_{en})^2 \leq 1 \hbox{ eV}~.
\eeqar{36}
For the RHC-mechanism, we use the presence of the two-neutrino matrix element
in the expression for $C_{\eta \eta}$ together with eq.~(32) to write
\beqar
\frac{1}{\tau^{0\nu}} = a^{-2}\left[\sum_n\lambda_nU_{en}V_{en}(m_n)^2\right]^2
                        \times \frac{4F_{33}^0}{F_{2\nu}}
                        \times \frac{1}{\tau^{2\nu}}~.
\eeqar{37}
The measured lifetime for two-neutrino decay in $^{76}$Ge is very close to
$10^{21}$ yrs and the ratio of phase space factors can be estimated from Table
A1 of Tomoda to be $3.3\times 10^3/\hbox{MeV}^4$. Taking the parameter $a$ to
be 0.5, we then obtain the bound
\beqar
\sum_n \lambda_n U_{en}V_{en}(m_n)^2  \leq 200 \hbox{ keV}^2
\eeqar{38}
If we include the interference terms between the two mechanisms, then we obtain
the usual quadratic forms in the two-dimensional lepton number nonconserving
parameter space.

\section{Implications of seeing no-neutrino decay}

Suppose now that at some future date, the no-neutrino mode will actually be
seen in $^{76}$Ge decay with a half-life of $2N^2\times 10^{24}$ yrs. What will
this mean for neutrino mass aside from the obvious implication that the mass
must be nonzero\Ref{23}?

The first implication is that the above inequalities will be replaced by one of
the equalities
\beqar
  \sum_n m_n \lambda_n (U_{en})^2 &= &\frac{1}{N} \hbox{ eV} \nonumber \\
  \sum_n \lambda_n U_{en}V_{en}(m_n)^2 &= &\frac{200}{N} \hbox{ keV}^2,
\eeqar{39}
depending upon which mechanism is at work. If we assume that all neutrino mass
eigenstates are much lighter than $q$, the momentum of the exchanged neutrino
(roughly 50 MeV), then we can use these equations to set lower limits on the
highest eigenvalue $m_{max}$.

In the case of the mass mechanism we find that
\beqar
 m_{max} \sum_n \lambda_n (U_{en})^2 \geq \frac{1}{N} \hbox{ eV}~;
\eeqar{40}
and, since the sum of squares of mixing coefficients $U_{en}$ can never exceed
unity, it follows that
\beqar
 m_{max}  \geq \frac{1}{N} \hbox{ eV}~.
\eeqar{41}
In the case of the RHC-mechanism, we find by a similar argument\Ref{14} that
\beqar
m_{max} \geq \frac{14}{\sqrt N} \hbox{ keV}~.
\eeqar{42}

The physical implications of these bounds are not insignificant, especially in
the case of the RHC-mechanism. Equation (42) implies that for no-neutrino
double beta decay to be observed, there must exist at least one neutrino with
a mass of several keV. Given the apparent demise of the 17 keV
neutrino,\Ref{24}
this  is unlikely; and thus RHC-induced double beta decay is also unlikely. The
mass-mechanism requires a neutrino of mass in the eV range or less, a more
plausible  possibility.

\section{Majoron emission}

One model of neutrino mass is based upon the coupling of neutrinos to a light
pseudo-Goldstone boson associated with the spontaneous breakdown of lepton
conservation.\Ref{11} This can give rise to double beta decay in which two
electrons  and a spinless boson, the Majoron, are emitted, but they are not
accompanied by  neutrinos. The half-life for this process can be expressed in
terms of the same  matrix element as occurs in no-neutrino decay, the coupling
$g_M$ of the  neutrino to the Majoron and a three-body phase space factor
$F_M$:
\beqar
 \frac{1}{\tau^{0\nu M}} = [(g_M) M^{0\nu}(0)]^2 F_M ,
\eeqar{43}
where $F_M$ is tabulated by Tomoda\Ref{17} and has a value of $1.3 \times
10^{-14}  \hbox{fm}^{-2} \hbox{yr}^{-1}$ for $^{76}$Ge.

We can search for the Majoron mode by studying the double beta decay spectrum
as a function of the sum $E$ of the electron energies. For the no-neutrino mode
$E$ is always equal to the total energy released, $Q$, because there are no
other leptons to carry off the available energy; for the two-neutrino mode, the
energy is shared between the electrons and neutrinos and so the spectrum is
continuous. It peaks below the mid-point $Q/2$, approaches the end-point like
$(Q - E)^5$ and is virtually zero for the last $20\%$ of the spectrum.

By contrast, the two electrons in the Majoron mode carry off most of the
available energy. Since the Majoron is spinless and the electrons have the same
helicity, they have to emerge predominantly in a back-to-back configuration in
order to conserve total angular momentum in the $0^+ \rightarrow 0^+$ nuclear
transition. Hence the Majoron tends to be emitted softly. As a result the
spectrum for the Majoron mode peaks in the region near the end-point where the
two-neutrino spectrum vanishes.

As we shall hear from Prof.~Moe,\Ref{25} there appear to be anomalous numbers
of
events  in this region in the observed spectra for $^{82}$Se, $^{100}$Mo and
$^{150}$Nd.  The fact that these nuclei have different $Q$ values tends to
suggest that the phenomenon is real, but it is too early to rule out some as
yet
unanticipated  background as the source of the anomaly. Recent experiments with
enriched  sources of $^{76}$Ge, which has a lower end-point than the other
nuclei, find no evidence for the anomaly\Ref{26} and argue against an
interpretation as the Majoron  mode. The effect Moe appears to observe could
still be real, but with a  completely different interpretation of a nuclear,
rather than particle nature.

\section{Limits on heavy neutrinos}

In addition to the light neutrinos that we have considered so far, we could
also discuss heavy neutrinos, with masses much greater than the mean neutrino
momentum of $q \approx 50 MeV$. For mass-induced no-neutrino decay, the crucial
factor in the leptonic matrix element is the product
\beqar
   P = \frac{qm_{\nu}}{q^2 + m_{\nu}^2}
\eeqar{44}
For light neutrinos, $m_{\nu} \ll q$ and
\beqar
   P \approx \frac{m_{\nu}}{q}~,
\eeqar{45}
while for heavy ones, $M_{\nu} \gg q$ and
\beqar
   P = \frac{q}{M_{\nu}}~.
\eeqar{46}
{}From the equivalence between these two forms for $P$, we obtain a `see-saw'
for
light and heavy neutrinos:
\beqar
  m_{\nu}M_{\nu} \approx q^2 \left( \frac{M_W}{M_{W'}} \right)^4
\eeqar{47}
where we have allowed for different gauge boson masses $M_W$ and $M_{W'}$
associated with the light and heavy neutrinos respectively. Taking $q$ to be 50
MeV and the ratio of boson masses to be about 1/10, we obtain
\beqar
m_{\nu}M_{\nu} \approx (m_e)^2.
\eeqar{48}
If the light mass is of order 1 eV, then the heavy one must be of order 100
GeV.

\section{Conclusion}

In the years since Cowan and Reines opened the era of {\em Experimental
Neutrino Physics}, great progress has been made in the study of double beta
decay. The two-neutrino mode has been observed in the laboratory with
half-lives as long as $10^{21}$ yrs and the so-called `geochemical method' has
been used to detect double beta decay with half-lives of order $10^{23}$
yrs.\Ref{27} Bounds on the no-neutrino mode have been extended beyond $10^{24}$
yrs\Ref{26} and, with the advent of enriched sources, it may be possible to
push the range of sensitivity to $10^{26}$ yrs, corresponding to an effective
mass of $0.1$ eV. We are still in search of the holy grail of double beta decay
without  neutrinos, and thus we are intrigued by the apparent anomaly found by
Moe.\Ref{25}  If it cannot be shown to be a background, then it should be
attacked experimentally on the same scale as the solar neutrino problem.

Our theoretical understanding of double beta decay has also advanced. We are
now able to calculate the nuclear matrix elements much more accurately than in
the early days,\Ref{19} and we have a much clearer understanding of the
{\em fundamental} role of neutrino mass in relation to the no-neutrino
mode.\Ref{14,\,23}  If this mode is eventually observed, then not only will we
be able to conclude that the neutrino has a mass, but we will also be able set
a lower bound on the mass of the heaviest mass eigenvalue. Such a bound would
be important for other phenomena that are associated with mass, for example
neutrino oscillations.

I hope that Fred will take much pleasure in the progress that has been and that
he will take paternal pride in Neutrino Physics and its practitioners.

\section*{References}
\begin{enumerate}
\frenchspacing
\item F. Reines in {\it Neutrino Physics and Astrophysics} (Proceedings of
      Neutrino '80) edited by E. Fiorini (Plenum Press, New York, 1982)
      p.11--28.
\item T. Kirsten, O. Schaeffer, E. Norton, and R.W. Stoenner, Phys. Rev. Lett.
      {\bf 20}, 1300 (1968).
\item S.R. Elliott, A.A. Hahn, and M.K. Moe, Phys. Rev. Letters {\bf 59}, 2020
      (1987).
\item C. L. Cowan, F. B. Harrison, L. M. Langer, and F. Reines, Nuovo Cim.
      {\bf 3}, 649 (1956).
\item S. L. Glashow, Nucl. Phys. {\bf22}, 579 (1961); S. Weinberg, Phys. Rev.
      Lett. {\bf 19}, 1264 (1967); A. Salam in
      {\it Elementary Particle Theory, Relativistic Groups, and Analyticity}
      edited by N. Svartholm (Almqvist and Wiksell, Stockholm,1968) p.~367.
\item E. Majorana, Nuovo Cimento {\bf 14}, 171 (1937).
\item G. Racah, Nuovo Cimento {\bf 14}, 322 (1937). The notion of
      neutrino-induced reactions was first discussed by H. Bethe and R.
Peierls,
      Nature {\bf 133}, 532 and 689 (1934).
\item R. Davis, Jr., Phys. Rev. {\bf 97}, 766 (1955).
\item B. Pontecorvo, Sov. Phys. JETP {\bf 6}, 429 and {\bf 7}, 1972 (1957). See
      also B. Pontecorvo in {\it Neutrino Physics and Astrophysics}
(Proceedings
      of Neutrino '80) edited by E. Fiorini (Plenum Press, New York, 1982)
p.52.
\item W. H. Furry, Phys. Rev. {\bf 56}, 1184 (1939).
\item Y. Chikashige, R.N. Mohapatra, and R.D. Peccei, Phys. Lett. {\bf 98B},
      265 (1981); G.B. Gelmini and M. Roncadelli, ibid {\bf 99B}, 411 (1981);
      and H. Georgi, S.L. Glashow, and S. Nussinov, Nucl. Phys. {\bf B193}, 297
      (1981).
\item M. Goeppert-Mayer, Phys. Rev. {\bf 48}, 512 (1935).
\item S. P. Rosen, Phys. Rev. {\bf D29}, 2535 (1984) and {\bf D30}, 1995 (E).
\item B. Kayser, S. Petcov, and S. P. Rosen (in preparation).
\item S. P. Rosen, in {\it Gauge Theories, Massive Neutrinos, and Proton Decay}
      edited by A. Perlmutter (Plenum Press NY 1981) p. 333.
\item J. Engel, W. C. Haxton, and P. Vogel, Preprint INT92-07-12 (submitted to
      Phys. Rev. C).
\item T. Tomoda , Rep. Prog. Phys. {\bf 54}, 53 (1991).
\item W. C. Haxton, Phys. Rev. Lett. {\bf 67}, 2431 (1991).
\item For a review, see talk by W. C. Haxton in Proceedings of Neutrino '92,
      preprint No.40427-21-N92, Institute for Nuclear Theory.
\item F. T. Avignone III, {\it et al.}, Phys. Lett. {\bf B256}, 559 (1991).
\item M. Moe, M. Nelson, M. Vient and S. Elliott, preprint UCI-NEUTRINO 92-1
      (1992).
\item A. Piepke, talk at XXVI International Conference on High Energy Physics,
      Dallas, Texas 5--12 August, 1992; D. Caldwell {\it et al.}, Nucl. Phys.
      (Proc. Suppl.) {\bf B13}, 547 (1990).
\item J. Schechter and J. W. F. Valle, Phys. Rev. {\bf D25}, 2951 (1982).
\item R. G. H. Robertson, Plenary talk at XXVI International Conference on High
      Energy Physics, Dallas, Texas 5--12 August, 1992.
\item M. Moe, following talk.
\item See A. Piepke, reference 22.
\item T. Bernatowicz {\it et al.}, Phys. Rev. Lett. (to be published) (1992).
\end{enumerate}

\end{document}